# DEVELOPMENT OF A μTPC DETECTOR AS A STANDARD INSTRUMENT FOR LOW ENERGY NEUTRON FIELD CHARACTERIZATION

D. Maire[1,2,*], J. Billard[2], G. Bosson[2], O. Bourrion[2], O. Guillaudin[2], J. Lamblin[2], L. Lebreton[1], F. Mayet[2], J. Médard[2], J.F. Muraz[2], J.P. Richer[2], Q. Riffard[2], and D. Santos[2]
[1] IRSN, 13115 Saint Paul-Lez-Durance, France
[2] LPSC (CNRS-IN2P3/UJF/INPG), 38000 Grenoble, FRANCE

**In order to measure energy and fluence of neutron fields, with energy ranging from 8 keV to 1 MeV, a new primary standard is being developed at the IRSN (Institute for Radioprotection and Nuclear Safety). This project, μ-TPC (Micro Time Projection Chamber), carried out in collaboration with the LPSC, is based on the nucleus recoil detector principle. The measurement strategy requires track reconstruction of recoiling nuclei down to a few keV, which can be achieved with a low pressure gaseous detector using a micro-pattern gaseous detector. A gas mixture, mainly isobutane, is used as a n-p converter to detect neutrons into the detection volume. Then electrons, coming from the ionization of the gas by the proton recoil, are collected by the pixelised anode (2D projection). A self-triggered electronics is able to perform the anode readout at a 50 MHz frequency in order to give the third dimension of the track. Then the scattering angle is deduced from this track using algorithms. The charge collection leads to the proton energy, taking into account the ionization quenching factor. This article emphasizes the neutron energy measurements of a monoenergetic neutron field produced at 127 keV. The measurements are compared to Monte Carlo simulations using realistic neutron fields and simulations of the detector response. The discrepancy between experiments and simulations is 5 keV mainly due to the calibration uncertainties of 10%.**

INTRODUCTION

In ionizing radiation field, facilities producing neutron fields are essential to study and to calibrate neutron detectors. To do so, neutron fields are characterized in energy and fluence by a standard spectrometer and then can be considered as references. The IRSN is associated with the French institute of metrology (LNE) for the French neutron references: the energetic distribution of the neutron fluence, the neutron fluence, the individual dose equivalent and kerma quantities in neutron metrology. To measure directly the energy distribution of neutron fields with energies below a few tens of keV, a new standard spectrometer, with a low energy threshold, is required at the Laboratory of Metrology and Neutron Dosimetry (IRSN/LMDN).

This project is undertaken in collaboration with the MIMAC team (LPSC/UJF/CNRS-IN2P3/INPG) which has developed the first prototype[2] for directional dark matter search [3]. The direct detection of dark matter is similar to the neutron measurements in the keV range because the interaction with matter of these particles induces in both cases nuclear recoils.

A NEW PRIMARY STANDARD SPECTROMETER

**Neutron energy measurement principle**

A nuclear recoil detector uses a converter to produce nucleus recoils of mass $m_A$ thanks to the elastic scattering of neutrons (mass $m_n$) onto these nuclei. The nucleus energy ($E_A$) measurement and the reconstruction of the scattering angle ($\theta_A$), enable to reconstruct directly the neutron energy ($E_n$), following the equation 1. This would allow the nuclear recoil detector to be a primary standard.

$$E_n = \frac{(m_n + m_A)^2}{4 m_n m_A} \times \frac{E_A}{\cos^2(\theta_A)} \quad (1)$$

If the nuclear recoil is a proton, $m_n \approx m_A$, the fraction of masses is approximately equal to 1. The maximum energy achievable for a recoiling nucleus is the highest for protons.

**Technical description of the μ-TPC**

The μ-TPC is a proton recoil detector and aims at characterizing low energy neutron fields [4], between 8 keV and 1 MeV. The use of a gas as a n-p converter

*Corresponding author: donovan.maire@irsn.fr

and the detection of proton recoils are the only answer to reach such an energy range.

The μ-TPC is divided in two zones: the conversion zone 17.7 cm in length and the amplification one, 256 μm in length of a bulk micromegas [5]. In the first zone, proton recoils stemmed from the neutron scattering lose a part of their kinetic energy by ionizing the gas producing a number of ion-electron pairs. A field cage surrounding the conversion zone produces a uniform electric field. This field enables to drift, toward the amplification zone, the electrons coming from the ionization. In the second zone a high electric field produces an avalanche which amplifies the signal up to a pixelised anode. The ions coming from this avalanche go up to the grid and the electrons go down to the anode. The charge collection on the grid leads to the ionization energy measurement.

The anode has an active area of 10.8x10.8 cm$^2$ and is segmented in pixels with a pitch of 424 μm. The 2D readout of the anode is performed by reading 256 strips in each dimension to access the X and Y positions. The pixelised anode is entirely read with a frequency of 50 MHz thanks to an efficient self-triggered electronics associated with a data acquisition system developed at LPSC [6][7]. The third dimension is therefore reconstructed by knowing the drift velocity of electrons in the conversion zone [8][9]. The drift velocity was estimated with the Monte Carlo code MAGBOLTZ [10] (i.e. 24.68 μm/ns). The scattering angle of the proton track is deduced from this 3D reconstruction.

The gas mixture used is: 60% $C_4H_{10}$ and 40% $CHF_3$ at 50 mbar. The $C_4H_{10}$ was chosen due to the high proportion of Hydrogen. The $CHF_3$ allows lowering the drift velocity to obtain more images of the tracks. The gas flow is provided by a gas control system dedicated to this detector. This system enables the pressure and the composition of the gas to be changed in order to adapt the converter to the neutron energy delivered. Each gas is filtered to remove impurities such as $O_2$ and $H_2O$ molecules.

**Proton energy calibration.**

To calibrate the detector, two X rays sources are used. The $^{109}$Cd and $^{55}$Fe sources produce respectively LX and KX rays with a mean energy of 3.04 keV and 5.92 keV. The X rays interact by photoelectric effect in the detector and produces photoelectrons with a lower energy due to the binding energy. An Auger electron is immediately emitted. The measured energy is then the X ray energy. The photoelectrons lose their kinetic energy by ionizing the gas and the secondary electrons are collected on the anode. Due to the transparence of the gas at high X ray energies, only X rays with energies lower than 10 keV can be used for the calibration.

The energies measured to calibrate the μ-TPC are very low compared to the maximum proton energy achievable (i.e. 127 keV). In addition, only two energies are used and their uncertainties are 0.2 keV for the $^{109}$Cd source and 0.04 keV for the $^{55}$Fe source according to the database ENDF B-VII.1. This calibration induced therefore an uncertainty of at least 12 keV at the maximum energy (i.e. 127 keV), assuming the linearity of the energy measurement.

**Ionization quenching factor.**

Protons lose only a part of its kinetic energy by ionizing the gas. In this way the measured energy is only the ionization energy, $E_{ion}$. To measure the initial proton energy, the Ionization Quenching Factor (IQF) need to be estimated. This factor is defined by the yield between the measured ionization energy of the nuclear recoil and the ionization energy of the electronic recoil with the same initial energy, $E_{initial}$ (equation 2) [9].

$$Q = \frac{E_{ion}}{E_{initial}^e} \qquad (2)$$

This factor depends on the gas mixture and the initial proton energy. The knowledge of this factor is required to measure low proton energies (i.e. $E_{initial}$ < 50 keV). This factor is calculated for this analysis with SRIM [11]. But previous studies, performed by the MIMAC team [12], have shown that SRIM calculations overestimate the IQF up to 20% of the total kinetic energy.

## MEASUREMENT OF A NEUTRON FIELD AT 127 KEV

**Initial proton energy measurement**

Taking into account the energy calibration, the nuclear recoil energy distribution is plotted on the figure 1.

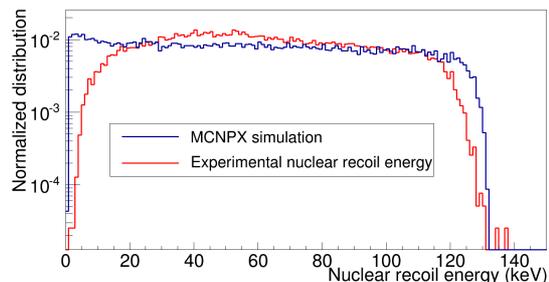

Fig. 1. Energy distributions obtained when a coincidence between strips of pixels X and Y on the anode is required (coincidence mode). The X ray calibration has been applied to obtain, from the measurement on the flash ADC, the energy in keV for each event. The blue curve corresponds to the MCNPX simulation.

To avoid boundary effects, events close to the edge of the detection zone are removed from the analysis. This cut enables also to remove events coming from the walls or the field cage of the detector.

At energies lower than 15 keV, the distribution decreases rapidly due to the detection threshold of the µ-TPC. A MCNPX simulation was done to compare the experimental nuclear recoil energy distribution to the expected proton energy distribution. The input neutron field was obtained with the TARGET code and is filtered by the solid angle of the system to reach a realistic neutron field, 127 keV with an FWHM of 7.2 keV. Then transport and conversion of neutrons are done in a realistic µ-TPC geometry. The experimental maximum proton energy, i.e. 130 keV, is 2 keV lower than the simulated one, but remains in the uncertainties given by the calibration process. The detection of photoelectrons due to gamma rays and the underestimation of scattered neutrons explain the difference between the distribution for an energy ranging between 30 keV and 70 keV.

**Initial proton recoil angle reconstruction**

The sampling of the pixelised anode every 20 ns gives a track (cloud of pixel) in three dimensions. Events are rejected of the analysis if the number of time sample is less than three.

The reconstruction method of the recoil angle includes a fit of a straight line to the cloud of pixels in the three spatial dimensions. This method using a linear fit is justified because of the very low deviation of proton during their recoil compared with the size of pixels at this energy range. This deviation has been calculated with the Monte Carlo code SRIM. The direction vector of the fitted line enables to calculate the angle between the track (fitted line) and the neutron incident direction. As the active area of the detector is wide (116 cm²) and close to the neutron source (72.5 cm), all neutrons are not parallel to the z-axis. The neutron direction vector is calculated by linking the neutron source position and the initial proton recoil position. The X and Y positions of each proton are calculated via the barycentre of the cloud of pixels. The Z position is unknown but it was fixed at the middle of the detection volume. A study with MCNPX simulations has shown this hypothesis does not modify significantly the neutron energy distribution. A scalar product between the direction vectors of the neutron and the proton gives the initial recoil angle.

**Reconstruction of the neutron energy**

Each nuclear recoil is supposed to be a proton because the hydrogen is the main component of the gas and its neutron scattering cross section is the highest for such energies compared to the other components of the gas. Since the ionization energy was measured, the proton IQF can be applied to calculate the initial proton energy. Once the initial proton energy and the initial proton recoil angle are measured event by event the neutron energy may be reconstructed via the equation 1. The figure 2 shows the agreement between the data and the equation 1, supposing every event is a proton recoil.

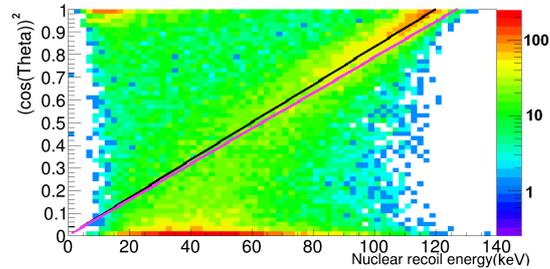

Fig. 2. Square of the cosine of the reconstructed recoil angle versus the nuclear recoil energy. The black and the pink curves correspond to the equation 1 for a proton recoil and respectively a neutron energy of 120 keV and 127 keV. The points represent the data obtained with a neutron field of an expected energy of 127 keV.

A part of the distribution plotted on the figure 2 follows the equation 1 calculated for a proton recoil and a neutron energy of 120 keV (black curve). The expected neutron energy is 127 keV. The difference comes mainly from the uncertainties on the energy calibration and the IQF overestimated by SRIM.

On the figure 2, three types of reconstructed events have been identified:

- Heavy nuclear recoils (i.e. carbon and fluorine) are wrongly analysed because their atomic masses and their IQF are supposed equal to the proton ones in the analysis. These recoils are located on the upper left side of the figure 2.
- Photoelectrons are still detected despite the selection of the coincidence of X and Y strips of pixels.
- Protons with an initial recoil angle higher than 40 degrees have energies less than 65 keV and the number of pixels fired decreases with the proton energy. The algorithm is then much less accurate to fit the right angle for angle higher than 40 degrees.

The experimental neutron energy distribution is shown on the figure 3 (green curve) and is compared to the theoretical neutron energy distribution (blue curve). Additionally the reconstruction algorithm was used to reconstruct the neutron energy taking into account the simulated detector response (red curve). This simulation, using MAGBOLTZ and SRIM calculations, is based on the model described in [13].

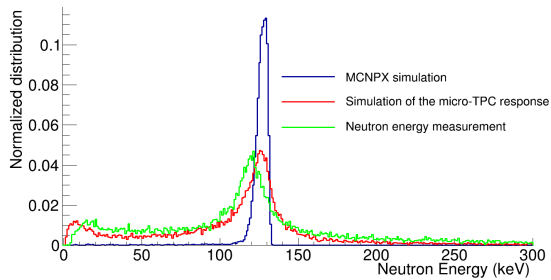

Fig. 3. Neutron energy distributions. The green curve corresponds to the experimental reconstructed energy distribution. The red curve is obtained by the simulation of the detector response. The neutron energy distribution simulated at 127 keV with MCNPX is also plotted (blue curve). The difference between the maximum of the experimental and simulated distributions is mainly due to the overestimated IQF given by SRIM and the uncertainties of the energy calibration.

Each distribution was normalized by the integral of the peak between 107 keV and 147 keV. The small bump at low energy is due to wrong reconstruction and heavy recoils, which have low energies due to their mass and their IQF. The peak is the one expected at 127 keV. The background of the distribution is due to incorrect reconstructions and neutron scattering on the walls. A cut at low energy is necessary for this analysis to removed events as heavy recoils. An optimized study of these events is required to go further.

Each distribution was fitted by a Gaussian function. The mean of the experimental distribution is 122 keV while the mean of the simulated distribution is 127.5 keV. The figure 2 has already shown this difference.

The resolution of the simulated distribution is calculated by the TARGET code with the target thickness and the kinematic calculations. The resolutions, defined as the FWHM over the mean energy, of the theoretical, simulated and experimental distributions are respectively 8%, 14% and 17%. The experimental resolution is higher, by a factor 1.2, than the resolution of the simulated distribution. This simulated resolution is slightly higher than the resolution given by previous simulations [4].

CONCLUSIONS

This experimental campaign demonstrates the ability of our system to reconstruct the energy of a monoenergetic neutron field at 127 keV.
The mean neutron energy found was 122 keV with an uncertainty of at least 12 keV mainly due to the calibration and IQF uncertainties. A new device will be soon available to improve the calibration and to measure the proton IQF [14].

The resolution (FWHM) of the experimental energy distribution is 17%, while the expected resolution is 14%, and only events with scattering angles lower than 40 degrees and initial recoil energies higher than 20 keV are quite well reconstructed. Then the reconstruction method will be improved by changing the fitting algorithms to take into account more events. Additionally a new chamber will be made to reduce the neutron scattering in the walls. The discrimination of particles thanks to the shape of the signal on the flash ADC is also planned to remove heavy nucleus from the analysis.